\documentclass{article}

\usepackage[a4paper]{geometry}
\usepackage{graphicx}
\usepackage[textwidth=8em,textsize=small]{todonotes}
\usepackage{amsmath}
\usepackage{natbib}
\usepackage{comment}
\usepackage{bm}
\usepackage{amsfonts}
\usepackage[hyphens]{url}

\begin{document}

\begin{center}
    \Large{A Fractional Model for Earthquakes}\\
    
  \normalsize Louis Davis$^{1,2}$, Boris Baeumer$^1$, Ting Wang$^1$\\
    $^1$Department of Mathematics and Statistics, University of Otago, Dunedin, New Zealand\\
    $^2$E-mail: \url{davislrs2000@gmail.com}
\end{center}

\begin{abstract}
This paper extends the existing fractional Hawkes process to better model mainshock-aftershock sequences of earthquakes. The fractional Hawkes process is a self-exciting point process model with temporal decay kernel being a Mittag-Leffler function. A maximum likelihood estimation scheme is developed and its consistency is checked. It is then compared to the ETAS model on three earthquake sequences in Southern California. The fractional Hawkes process performs favourably against the ETAS model. Additionally, two parameters in the fractional Hawkes process may have a fixed geophysical meaning dependent on the study zone and the stage of the seismic cycle the zone is in.
\end{abstract}

\section{Introduction}
\label{sec:intro}

Self-exciting point process models have been used to model earthquakes since the works of \cite{hawkes1973cluster}. Since then, the \textit{Epidemic type aftershock sequence} (ETAS) model introduced by \cite{Ogata1988} has enjoyed much success, however it still has shortcomings. Specifically, its long-range forecasting abilities are limited due to its simple background rate and its analytic intractability \citep[e.g.][]{harte2013bias}. 

Recently, \cite{chen2021fractional} and \cite{https://doi.org/10.48550/arxiv.2211.02583} introduced and studied a so called ``fractional Hawkes process". It was noted that the fractional Hawkes process is of the same type as the ETAS model due to the asymptotic behaviour of the kernel function behaving as a power law. The kernel functions have other differences as well. The kernel function of the fractional Hawkes process has an integrable singularity at $0$ and is analytically tractable due to its closed form Laplace transform and connection to fractional calculus. This allowed \cite{chen2021fractional} to investigate spectral properties and find a closed form expected intensity with relative ease. 

In contrast, the ETAS model is relatively hard to work with analytically, and its kernel function is always bounded. We expect this to mean that the fractional Hawkes process will effectively model strongly clustered data since for any parameter value the conditional intensity function becomes infinite. The ETAS model is unable to do so without forcing a parameter value towards infinity. Moreover, the analytic tractability of the fractional Hawkes process should allow suitable model extensions to be performed. Such extensions would relax the assumption of event time and magnitude independence potentially allowing longer-range forecasts to be done than is currently possible with the ETAS model. Furthermore, as will be discussed in Section \ref{sec:Model}, the fractional Hawkes process has a direct link to the Fractional Zener model studied by \cite{metzler2003fractional} which describes relaxation of viscoelastic materials and so the model has both an empirical and scientific backing. This basis suggests that such a model is worthy of investigation, after some slight modification.

The objective of this paper is to modify the  fractional Hawkes process to better describe earthquake behaviour and then compare its performance to the ETAS model. The fractional Hawkes process is introduced and modified in Section \ref{sec:Model}. A numerical maximum likelihood estimation scheme is discussed in Section \ref{sec:ParamEst}. We then apply the fractional Hawkes process to aftershock sequences from Southern California and compare its performance against that of the ETAS model. Residual analysis of both models is performed to check model fit. The fractional Hawkes process is then modified, by removing its background rate term, and refit to the data in Section \ref{sec:RemoveLamb0}. This is to investigate whether the scientific backing of the model is still justified. Finally, future developments are discussed in Section \ref{sec:Conc}. 

\section{The Fractional Hawkes Process}\label{sec:Model}
 \cite{Hawkes1971} formulated a self-exciting point process, since called the Hawkes process. It is a point process with conditional intensity function
\begin{equation}\label{eq:HPint}
    \lambda(t|H_t)=\lambda_0+\alpha \sum_{t_i<t} f(t-t_i),
\end{equation}
where $H_t$ is the history, $\lambda_0 > 0$ the arrival rate of the background or ``immigrant" events, $\alpha>0$ the parameter controlling the overall intensity, and $f(t) \geq 0$ an integrable function with support on $\mathbb{R}^+$. As an aside, $f(t)$ will be referred to as the ``kernel" function. To now construct the fractional Hawkes process define an alternative kernel function

\begin{equation}\label{eq:fbeta}
    f_\beta(t):=-\frac{d}{dt}E_{\beta,1}(-t^\beta)=t^{\beta-1}E_{\beta,\beta}(-t^\beta),
\end{equation}
where $E_{\beta,\beta}(-t^\beta)$ is the two variable Mittag-Leffler function defined by the power series
\begin{equation}
E_{\gamma,\delta}(z):=\sum_{n=0}^\infty \frac{z^n}{\Gamma(n\gamma+\delta)}.  
\end{equation}
The corresponding one variable Mittag-Leffler function being $E_{\beta,1}(z)$. Let $z=-t^\beta$, then for $t \geq 0$, $E_{\beta,1}(-t^\beta)$ interpolates between a stretched exponential for small $t$ and a power law of index $\beta$ for large $t$. This is the survival function of a Mittag-Leffler random variable $X$, with infinite mean, and has corresponding density function $f_\beta(t)$. The fractional Hawkes process formulated by \cite{chen2021fractional} and \cite{https://doi.org/10.48550/arxiv.2211.02583} has the kernel $f_{\beta}(t)$.

We make two changes to the fractional Hawkes process to increase its suitability to model earthquakes. The first is to include an exponential weighting to the kernel function. The ETAS model from the ``PtProcess" R package developed by \cite{Harte2010} has conditional intensity function
\begin{equation}\label{eq:ETAS}
    \lambda(t|H_t)=\mu+A\sum_{t_i<t} e^{\delta(M_i-M_0)}\left(1+\frac{t-t_i}{c_E} \right)^{-p}.
\end{equation}
We add the $e^{\delta(M_i-M_0)}$ term to the fractional Hawkes process. This term is an exponential weighting, which means greater magnitude earthquakes increase the conditional intensity function more. This is based on the approximate exponential distribution of magnitudes as seen in \cite{Ogata1988} when the ETAS model was first studied.

The second modification is the introduction of an extra parameter to allow for a time-scale, somewhat analogous to the $c_E$ parameter in the ETAS model given by equation \eqref{eq:ETAS}, and this is where the link to the fractional Zener model begins. Early studies, such as \cite{nutting1921new} and \cite{gemant1936method}, in viscoelastic materials have suggested that relaxation phenomena follows a form of power law. Subsequent studies, examples being \cite{caputo1969elasticita,caputo1976vibrations}, used models with fractional derivatives for geological strata. Elastic rebound theory, developed by \cite{Reid1911}, is the chief explanation to how earthquakes occur. The theory requires the earth's crust to be some viscoelastic material and so models of such materials are worth considering in order to better describe earthquake behaviour.  Recently, \cite{metzler2003fractional} developed a more direct link between these materials and the Mittag-Leffler function. They extended the Zener model, to the so called  ``fractional Zener model." The model is derived by considering a Fractional Fokker-Planck equation (FPPE)
\begin{equation}\label{eq:FPPE}
    \frac{\partial W(\bm x,t)}{\partial t}=\mathbb{D}_t^{1-\beta}\left(\nabla\cdot\frac{V'(\bm x)}{m\eta_\beta}+K_\beta \nabla^2\right)W(\bm x,t),
\end{equation}
 where $\mathbb{D}_t^{1-\beta}$ is the Riemann-Liouville fractional derivative defined as
 \begin{equation}
     \mathbb{D}_t^{1-\beta} f(t) =  \frac{1}{\Gamma(\beta)} \frac{d}{dt} \int_0^\infty f(t-x)x^{\beta-1}dx.
 \end{equation}
 $K_\beta$ is the generalised diffusion coefficient, $\eta_\beta$ is the generalised friction and $V(\bm x)=-\int^{\bm x}_aF(\bm s)d\bm s$ is some external potential field acting on a test particle. 

Take the potential field to be constant, and replace the density function with the strain field $\varepsilon(\bm x,t)$ where $\bm x$ is the spatial variable and $t$ time. Equation (\ref{eq:FPPE}) becomes
\begin{equation}\label{eq:FracZener}
    \frac{\partial \varepsilon}{\partial t}(\bm x,t)=\mathbb{D}_t^{1-\beta}K_\beta \nabla^2\varepsilon(\bm x,t).
\end{equation}
One finds that the equation governing the mode relaxation, i.e. determining the temporal change of the strain modes, $\varepsilon(\bm k,t)$, can be found through a separation of variables ansatz  \citep[e.g.][]{metzler2000random}. Assuming $\varepsilon(\bm x,t)=T(t)\Psi(\bm x)$ then $T(t)$ satisfies the fractional differential equation
\begin{equation}\label{eq:temporalRelaxation}
    \frac{d T}{dt}=-|\bm k|^2\mathbb{D}_t^{1-\beta}T(t).
\end{equation}
The solution of equation (\ref{eq:temporalRelaxation}) is then given in terms of the Mittag-Leffler function
\begin{equation}\label{eq:MLFFracZener}
    T(t)=E_{\beta,1}(-K_\beta |\bm k|^2t^\beta).
\end{equation}
 The temporal change in the strain field is given by the first derivative of $T(t)$. Specifically
\begin{equation}
    \frac{dT}{dt}=-K_\beta |\bm k|^2 t^{\beta-1}E_{\beta,\beta}(-K_\beta |\bm k|^2t^\beta),
\end{equation}
for $t>0$ this is negative and hence the strain field monotonically $\varepsilon(\bm x ,t) \to 0$ as $t \to \infty$. 

The parameter $K_\beta$ and the spatial modes are unknown for a general geological region. Furthermore, since the lowest spatial eigenmode persists for the longest it dominates the relaxation behaviour of the viscoelastic material.

Let $c^\beta = K_\beta|\bm k|^2$. It is then natural to select
\begin{equation}
    cf_\beta(ct)=-\frac{dT}{dt}=c^\beta t^{\beta-1}E_{\beta,\beta}(-(ct)^\beta),
\end{equation}
as the kernel function for the model since then the temporal decay of the conditional intensity function is the same as the temporal decay of the strain field $\varepsilon(\bm x,t)$. It is worth noting that $cf_\beta(ct)$ has an integrable singularity at $t=0$. This feature should be what allows the model to best describe clustered mainshock-aftershock sequences. Therefore, the modified fractional Hawkes process for earthquakes has conditional intensity function
\begin{equation}\label{eq:FHPint}
    \lambda(t|H_t)=\lambda_0+\alpha \sum_{t_i<t}e^{\gamma(M_i-M_0)}cf_\beta(c(t-t_i)).
\end{equation}

Theoretically, $c^\beta = K_\beta|\bm k|^2$ are parameters in a physical model. One would expect they have a fixed value dependent on certain physical and geological properties of the studied region and so independent of the specific aftershock sequence studied.

\section{Parameter Estimation and Residual Analysis}
\label{sec:ParamEst}
To compare the ETAS model and fractional Hawkes process three mainshock-aftershock sequences from the Southern California seismological zone were selected and studied. Like when the ETAS model was first studied by \cite{Ogata1988} we assume independence of event time and magnitude distributions to allow for easier comparison between the ETAS and fractional Hawkes process models. Hence, given the data $\{(t_i,M_i)\}_{1\leq i \leq N}$, where $t_i$ is the occurrence time of the $i^{th}$ earthquake and $M_i$ its corresponding magnitude, we will maximise the conditional log-likelihood, which is
\begin{equation}\label{eq:loglik}
    \ell(\bm \theta)=\log(L(\bm \theta))=\sum_{i=1}^N \log(\lambda(t_i;\bm \theta))-\int_0^T \lambda(t;\bm \theta)dt,
\end{equation}
where $\bm \theta=(\lambda_0,\alpha,\gamma,\beta,c)$ and $\lambda_0,\gamma,c \geq 0$ with $\alpha,\beta \in [0,1]$ for the fractional Hawkes process. We are only concerned with the conditional log-likelihood since we assume the magnitude distribution of both the fractional Hawkes process and ETAS models are the same exponential distribution. This comes from the assumption that the event time and magnitude are independent, hence we can separately estimate parameters in the conditional intensity function and magnitude distribution. Since we want to investigate the performance of the fractional Hawkes process against the ETAS model, specifying different independent magnitude distributions would only make comparison of the models more difficult.

Define
\begin{equation}\label{eq:comp}
    \Lambda(T)=\int_0^T \lambda(t;\bm \theta)dt,
\end{equation}
then by substituting equation \eqref{eq:FHPint} into equation \eqref{eq:comp} and performing the integration we have
\begin{equation}
    \Lambda(T)=\lambda_0T-\alpha \sum_{i=1}^{N} e^{\gamma(M_i-M_0)}[E_\beta(-(c(T-t_i))^\beta)-1],
\end{equation}
and hence
\begin{multline}\label{eq:FHPloglik}
    \ell(\bm \theta)=\sum_{i=1}^{N} \log\left(\lambda_0+\alpha  \sum_{t_k<t_i} e^{\gamma(M_k-M_0)}cf_\beta (c(t_i-t_k)) \right) \\ -\lambda_0T+\alpha \sum_{i=1}^{N} e^{\gamma(M_i-M_0)}[E_{\beta,1}(-(c(T-t_i))^\beta)-1].
\end{multline}

\subsection{Likelihood Estimation and Consistency}
Equation \eqref{eq:FHPloglik} was then maximised numerically in MATLAB. The Mittag-Leffler function can be computed using the ``ml.m" function obtained from the MATLAB Central File Exchange by \cite{Garrappa2022A}. This function uses a sophisticated optimal parabolic contour algorithm developed by \cite{garrappa2015numerical} which had parts of it translated into C++ for more efficient computation. 

One would expect the conditional log-likelihood function to have the same asymptotic properties as the likelihood function under standard conditions such as stationarity and ergodicity of the joint series $\{(t_i,M_i)\}_{1\leq i \leq N}$. However to confirm this, an algorithm was implemented to simulate synthetic data from the fractional Hawkes process and then estimate parameters from this synthetic data. 

\begin{figure}
    \centering
    \includegraphics[width=\textwidth]{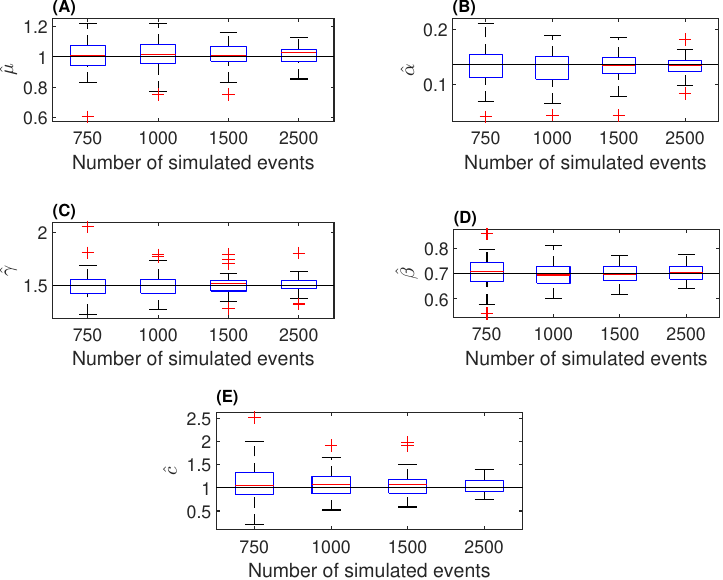}
    \caption{Boxplots (A) $\hat \lambda_0$, (B) $\hat \alpha$, (C) $\hat \gamma$, (D) $\hat \beta$ and (E) $\hat c$ with true parameter values $\alpha = e^{-1}$, $\lambda_0=1$, $\beta=0.7$, $\gamma=1.5$ and $c=1$. The true value is given by the solid black line.}
    \label{fig:Consistency}
\end{figure}

The synthetic data was generated by using the thinning method (e.g. \citealp{lewis1979simulation}, \citealp{ogata1981lewis}). The magnitude of each event was simulated using i.i.d tapered Pareto random variables as suggested by the ``PtProcess" R package \citep[e.g.][]{Harte2010}. The tapered Pareto is the minimum of a exponential random variable with parameter $\eta^{-1}$, increased by $M_0$, and a generalised Pareto random variable with shape parameter $\xi^{-1}$, threshold parameter $M_0$ and scale parameter $M_0\xi$. The parameter values used were $\eta=2$, $\xi = 3$ and $M_0=2.5$. We simulate 50 data sets using the same set of parameters for each of the sample sizes: 750, 1000, 1500 and 2500. We then estimate the parameters by numerically maximising the log-likelihood function in equation \eqref{eq:FHPloglik}. The boxplots of the estimated parameters are in Figure \ref{fig:Consistency}. Removing trials with obvious underflow or overflow errors it is clear the mean is centered around the true parameter value with decreasing variance suggesting consistency.

\subsection{Data Selection and Model Comparison}
The three data sets selected are named ``JT",``LM" and ``HM" corresponding to the events following three significant earthquakes and corresponding to their own seismic cycle. Summary information of the three catalogues is contained in Table \ref{tab:Geographicinfo}. The raw data was collected from the United States Geological Survey (USGS) website last accessed on 13/4/22. In all cases, the time window is the time of the first event to last event to eliminate the arbitrary nature of selecting the end point of the time interval considered.

\begin{table}
\caption{\label{tab:Geographicinfo} Summary information of the data sets, where $N$ is the size of the data set, $M_0$ the minimum magnitude for completeness as determined by the G-R law, times are in UTC and $M_1$ is the magnitude of the first event.}
    \centering
    \begin{tabular}{|c c c c c c c|}
    \hline
Catalogue & $N$ & $M_0$ & Time of first event&Time of last event &$M_1$ & Epicentre  \\
     
    JT  &  761 & 2.5 &23/04/92 04:50:23&28/06/92 05:48:05 &6.1& 33.96N, 116.32W\\ 
    LM & 1348 & 2.75 &28/6/92 11:57:34&31/12/92 07:49:10&7.3& 34.20N, 116.44W \\ 
    HM & 1023 & 2.5 & 16/10/99 09:46:44&21/1/00 17:08:00 & 7.1& 34.60N, 116.27W \\
      \hline 
    \end{tabular}

\end{table}

\begin{figure}
    \centering
    \includegraphics[width=\textwidth,trim={0.5cm 5cm 0.5cm 5cm}]{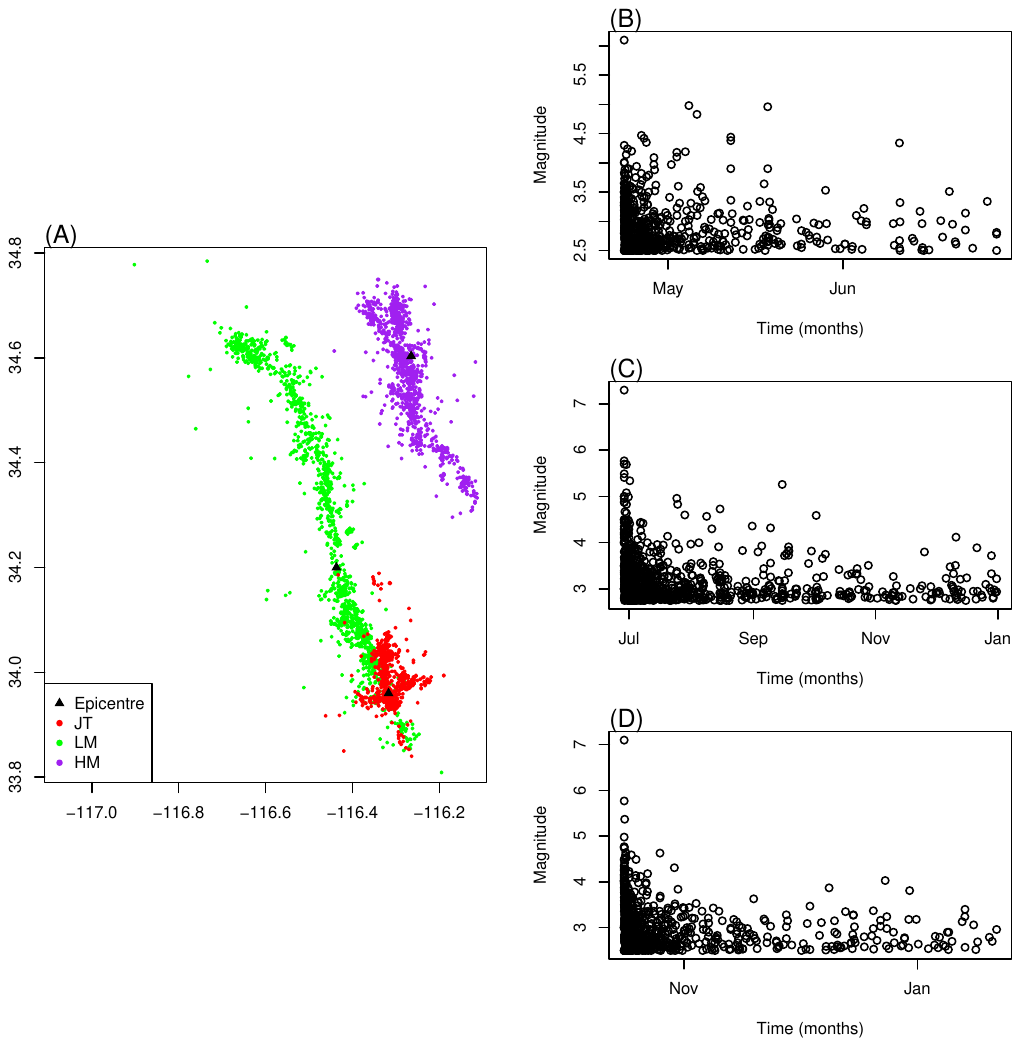}
    \caption{(A) Occurrence map of the studied catalogues. (B) Magnitude versus time plot for the ``JT" catalogue. (C) Magnitude versus time plot for the ``LM" catalogue. (D) Magnitude versus time plot for the ``HM" catalogue.}
    \label{fig:MAP}
\end{figure}
Each data set only contains events along the same fault line within the rectangular region with latitudes between $33.8$N and $34.8$N and longitudes between $117.1$W and $116.1$W as in Figure \ref{fig:MAP} (A). These data sets were selected for two reasons. Firstly, the restrictive geological setting should give the best possible chance for $c^\beta$ to be constant, hence providing evidence to support the scientific basis of the fractional Hawkes process without considering time or space varying $c^\beta$. Secondly, the magnitude threshold $M_0$, (selected so that the events above this cutoff obey the Gutenberg-Richter(G-R) law) is low for all three catalogues, providing large enough sample sizes for the analysis and model comparison.

Both models were then fitted to the data sets with 30 random initial values in the set $[0,1]^5$. The fits of the ETAS model was done by using a maximum likelihood estimator from the ``PtProcess" package in R by \cite{Harte2010}.  Doing so yields estimates given in Tables \ref{tab:FHPestimates} and \ref{tab:ETAS}.
 
 \begin{table}
 \caption{\label{tab:FHPestimates} MLEs and the negative log-likelihood for the Fractional Hawkes Process.}
 \centering
\begin{tabular}
{ |p{1.7cm} p{1cm} p{1cm} p{1cm} p{1cm} p{1cm} p{1cm} p{1.4cm}| }
 \hline
 Catalogue& $\hat \lambda_0$ &$\hat \alpha$&$\hat \gamma$ & $\hat \beta$& $\hat c$ &$\hat{c}^{\hat \beta}$&$-\ell(\hat{\bm \theta})$\\
 
JT &0.671 &0.495 & 1.178 &0.573 &6.946&2.920&-2302.0 \\

LM &0.414&0.402&1.273&0.665&0.849&0.707&-3325.5\\

HM &0.900 &0.180 &1.634&0.759&0.879&0.907&-3232.0\\
 
\hline
\end{tabular}

\end{table}

\begin{table}
\caption{\label{tab:ETAS}MLEs and the negative log-likelihood for the ETAS model.}
\centering
\begin{tabular}{ |p{1.7cm} p{1cm} p{1cm} p{1cm} p{1.05cm} p{1cm} p{1.5cm}| }
 \hline

 Catalogue& $\hat \mu$ &$\hat A$&$\hat\delta$ & $\hat c_E$ & $\hat p$&$-\ell(\bm \theta)$\\

JT &0.203&34.717&1.065&0.004&1.185&-2301.6\\

LM &0.345&0.917&1.249&0.255&1.552&-3311.3\\

HM &0.944&0.021&2.228&0.960&1.760&-3234.8\\ 

KK & 0.664& 2.863 & 0.813 & 0.087 & 1.416& -8840.2\\\hline

\end{tabular}

\end{table}
 
Comparison of the fractional Hawkes process and ETAS models is done using the Akaike Information Criterion, AIC \citep{akaike1974new}. As is standard, $\mbox{AIC} = -2\ell(\hat{\bm \theta}|\bm x)+2k$ where $\ell(\hat{\bm \theta}|\bm x)$ is the maximum log-likelihood and $k$ is the number of parameters in the model. A smaller AIC value suggests a more parsimonious model for the data, and as a rule of thumb, a difference of two or more is considered significant. Since AIC is widely used for comparing earthquake models with a low number of parameters, it is the appropriate information criteria to use for model selection. Define $\Delta$AIC = FHP AIC-ETAS AIC. Relevant values are in Table \ref{tab:AICFHPvETAS}.
 
\begin{table}
\caption{\label{tab:AICFHPvETAS}AIC scores for the differing models for each catalogue.}
\centering
\begin{tabular}{ |p{2cm} p{2cm} p{2cm} p{1.5cm}| }
 \hline

 Catalogue& FHP AIC& ETAS AIC & $\Delta \mbox{AIC}$ \\

 JT &-4594.0&-4593.2&-0.8 \\

 LM & -6641.1&-6612.6&-28.5\\

 HM &-6454.0&-6459.6 &5.6\\\hline
 
\end{tabular}

\end{table}
 
 The log-likelihoods and AIC scores indicate the fractional Hawkes process captures more information than the ETAS model at a non-significant level on the ``JT" data set, and at a significant level on the ``LM" data set. The ETAS model out performs the fractional Hawkes process at a significant level on the ``HM" data set.
\subsection{Residual Analysis}
While AIC provides evidence to which model explains the data the best, it does not rule out the existence of a better model that has not been considered. We therefore implement two of the graphical tests used in \cite{Ogata1988} to determine whether the ETAS model and fractional Hawkes process explain the main features of the data. For a set of data $\{t_i\}_{1 \leq i \leq N}$ generated from a point process with conditional intensity $\lambda(t)$, the sequence $\{\tau_i\}_{1 \leq i \leq N}$, known as the residual process, is defined by
\begin{equation}
    \tau_i=\int_0^{t_i}\lambda(t)dt.
\end{equation}
The residual process, calculated using the estimated conditional intensity $\hat \lambda(t)$, should be a Poisson process with rate $1$ if the model fits the data well \citep{Papangelou1972}. We then plot the mean removed transformed time sequence against the x-axis, and include the $95$\% error bounds of the Kolmogorov-Smirnov test statistic calculated using the ``sfsmisc" CRAN package \citep{Maechler2022}.

Since $\{\tau_k\}_{1 \leq k \leq N}$ should be a Poisson process, the sequence $X_k=\tau_k-\tau_{k-1}$ ought to be a sequence of i.i.d exponentially distributed random variables with parameter $1$. We then transform these to i.i.d uniform random variables on $[0,1)$ by $U_k=1-\exp\{-X_k\}$ and we perform a graphical test for serial correlation by plotting $U_k$ against $U_{k+1}$. 

\begin{figure}[!ht]
    \centering
    \includegraphics[width=0.65\textwidth]{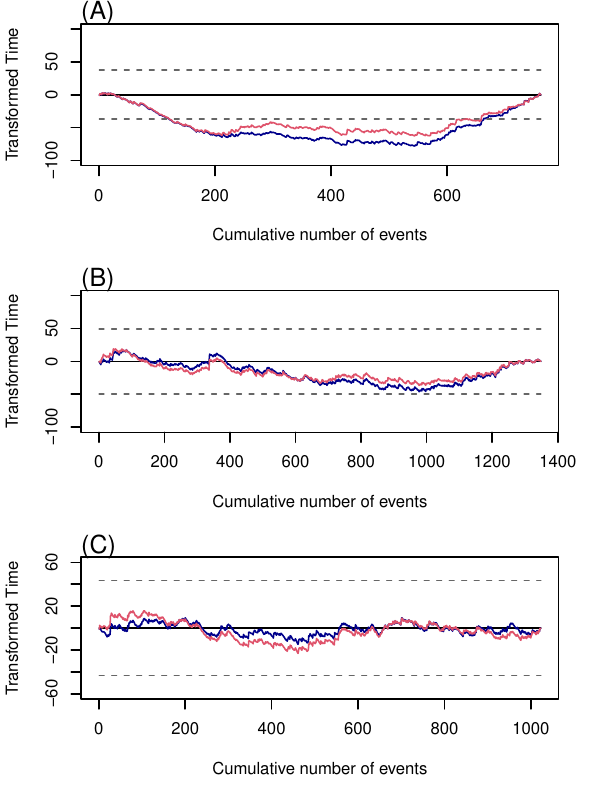}
    \caption{Residual analysis. (A) corresponds to ``JT", (B) to ``LM" and (C) ``HM". For all plots the red curve corresponds to the fractional Hawkes process, while the blue curve corresponds to the ETAS model. Dashed lines are the 95\% confidence bounds.}
    \label{fig:TotalRCT}
\end{figure}

Figure \ref{fig:TotalRCT} indicates that both models capture most features of the ``LM" and ``HM" data sets. In general, both models underpredict the total number of events, but within the permissible error bounds. Both models underpredict the number of events in the ``JT" catalogue for the majority of its duration. This can be explained by the delayed onset of quiescence. \cite{ogata2003and} and \cite{marsan2005methods} noted that seismicity was normal up until about event 200 before quiescence began, which is when both models majorly underpredict the total number of events. Beyond event 200, both models do not further underpredict the number of events. The quiescent period is influencing the model in a way so that in the immediate aftermath of the main event the models underestimate the total number of events to have any chance of explaining the rest of the catalogue. This influence is significant enough that it causes the models to fail the residual checks.  No concern of serial correlation was found between $U_k$ and $U_{k+1}$ for all three data sets as seen in Figure \ref{fig:TotalCorrelation} in Appendix A.

Furthermore, Figure \ref{fig:TotalRCT} can be used to explain the differences in AIC values for both models. For all data sets and both models, the final residual $\tau_N \approx N$, where $N$ is the size of the data set. Recalling equation \eqref{eq:loglik}
\begin{equation*}
\ell(\bm \theta)=\sum_{i=1}^N \log(\lambda(t_i;\bm \theta))-\int_0^T \lambda(t;\bm \theta)dt,
\end{equation*}
and using that $T=t_N$ for all data sets, the difference in log-likelihood is entirely due to the differing intensities at the event times. Models with a greater sum of log intensity will therefore have the larger log-likelihood and hence the AIC values will be in favour of that model in the case they have the same number of parameters.  Since $f_\beta(ct)$ has an integrable singularity at $t=0$ the fractional Hawkes process should be able to better model clustering as for any parameter value the intensity becomes arbitrarily large. The ETAS model however, requires $A$ to become increasingly large since the largest the kernel of the model can be is $A$ (when $t=t_i$ and momentarily ignoring magnitude). However, this highlights a potential weakness of the fractional Hawkes process in that the intensity always has an unbounded right limit when approaching an event time. Therefore, the fractional Hawkes process ought to better model shorter term data sets with strong clustering of events. Asymptotically $f_\beta$ decays a power law of index $\beta+1$. For all data sets $1+\hat \beta>\hat p$, which implies the intensity of the fractional Hawkes process decays faster than that of the ETAS model. Therefore, if there are many long periods with few events, the ETAS model should out perform the fractional Hawkes process because the sum in equation \eqref{eq:loglik} will be less in the fractional Hawkes process log-likelihood than the ETAS log-likelihood. 

Referring to Figure \ref{fig:MAP} (C) and (D) there is more clustering of events for a longer period of time in the ``LM" data set when compared to the ``HM" data set. Consequently, the intensity of the fractional Hawkes process at event times is generally greater than the ETAS model explaining the better fit to the ``LM" data set. We tested this theory on a sequence of events following the magnitude 7.8  Kaik\=oura earthquake of 2016 in New Zealand with data obtained from GeoNet Quake search which has strong clustering. The data set has 2398 events between 11:02:56 13-11-16 and 23:26:14 31-12-16 UTC. The study region was rectangular with latitudes ranging between $-43.5^{\circ}$ S and $-42^{\circ}$ S and longitudes between $171.59^{\circ}$ E and $174.45 ^{\circ}$ E. The minimum cutoff magnitude $M_0=2.75$. The fractional Hawkes process has a AIC of -17677.27, and the ETAS model has an AIC of -17670.41. Furthermore, both models pass the residual checks providing more evidence for our theory.

Each parameter estimate given in Tables \ref{tab:FHPestimates} and \ref{tab:ETAS}  has a physical interpretation. For the ``JT" and ``LM" data sets, the background rate ($\lambda_0$ and $\mu$ respectively) of the fractional Hawkes process is greater than that of the ETAS model. Furthermore, for these two data sets the fractional Hawkes process tends to trigger more immigrants than the ETAS model since $\hat \gamma>\hat \delta$. The estimates $\hat \alpha$ and $\hat A$ control the overall intensity of the process, and hence the total number of events. For the ETAS model the $i^{th}$ event increases the conditional intensity (in the right limit as $t \to t_i^+$) by $\hat Ae^{\hat \delta (M_i-M_0)}$. Therefore, $\hat A$ directly controls how great the estimated conditional intensity can be at any given time. 

Of most importance, to the fractional Hawkes process, is that the value $\hat c^{\hat \beta}$ differs between each of the data sets casting doubt on its hypothesised fixed value for this small study zone. However, we investigate the physical interpretation of $\hat c^{\hat \beta}$ by using the 95\% and 99\% asymptotic confidence intervals for $\hat \beta$ and $\hat c$ to calculate the most extreme value that $\hat{c}^{\hat \beta}$ could take. Calculating these, assuming a sufficient sample size for normality, with the numerically estimated Hessian matrix one obtains the values given in Table \ref{tab:cbetaCI}.

\begin{table}
\caption{\label{tab:cbetaCI}Most extreme values of $\hat c^{\hat \beta}$ using likelihood confidence intervals.}
\centering
\begin{tabular}{ |p{1.7cm} p{1.7cm} p{1.7cm} p{1.7cm} p{1.7cm}| }

 \hline
 Catalogue& Min 95\%&Max 95\%&Min 99\%& Max 99\% \\

JT &2.635&3.510&2.523&3.678\\

LM &0.598&1.143&0.487&1.228\\

HM &0.789&1.009&0.7481&1.043\\ \hline

\end{tabular}

\end{table}
 
It is entirely possible that the value of $c^\beta$ is the same for the ``LM" and ``HM" data sets, while it is clearly different for the ``JT" data set. This could be indicative that $c^\beta$ depends not only the geology of the region, as first suspected, but also which stage of the seismic cycle the region is in. The ``JT" data set is a quiescent period of this study zone, and since $c^\beta \propto K_\beta$, it is possible the generalised diffusion coefficient changes with time. However, since the fractional Hawkes process does not pass the graphical residual test on the ``JT" data set, the evidence to support this conclusion is weak at best.

\section{Removing Background Rate}\label{sec:RemoveLamb0}
\begin{figure}[!ht]
    \centering    \includegraphics[width=\textwidth]{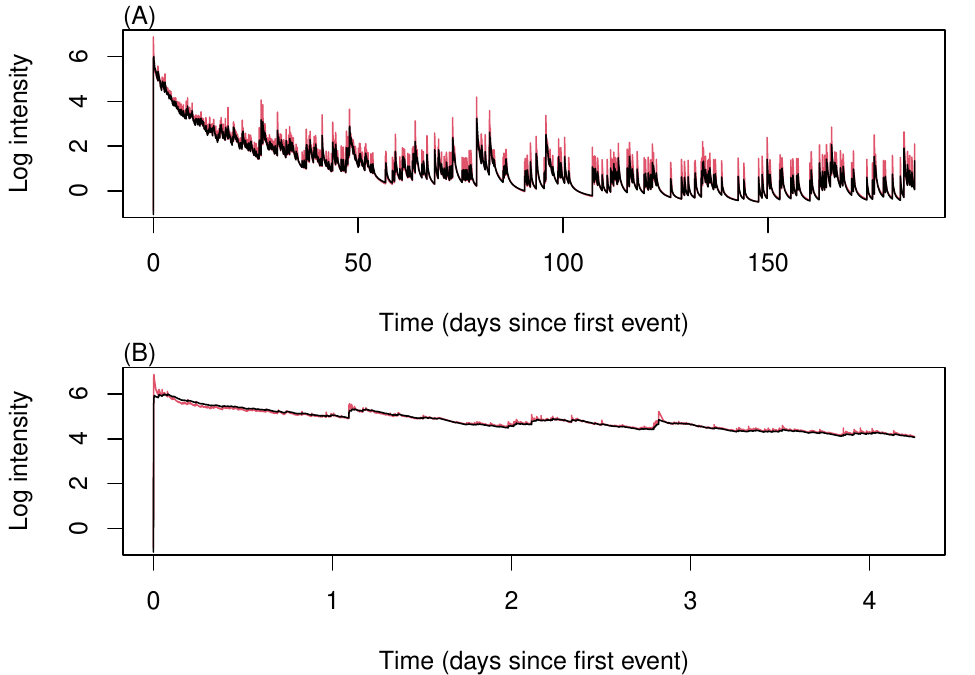}
    \caption{Log intensity for the ETAS (black) and fractional Hawkes process (red) on the ``LM" data set. (A) is for the duration of the catalogue while (B) is for the first 600 events.}
    \label{fig:LogInt}
\end{figure}

While both the fractional Hawkes process and ETAS models were successful in modelling the ``LM" and  ``HM" catalogues, they failed on the ``JT" catalogue. It has been seen that this period is relatively quiescent (e.g. \citealp{ogata2003and} and \citealp{marsan2005methods}). In our case, both models breach the lower confidence interval for our transformed time plot. Therefore, the estimated intensity is less than would be expected from reality. Referring to Tables \ref{tab:FHPestimates} and \ref{tab:ETAS} for all three data sets the estimated background rate seems relatively high for both models. There is a possibility that the models are over-predicting the number of background events, potentially distorting the estimates of $\alpha,\gamma,\beta$ and $c$ for the fractional Hawkes process which are the parameters attempting to explain the nontrivial clustering of events. Furthermore, for a significant length of time, the estimated intensity is much greater than $\hat \lambda_0$, which therefore contributes little to the estimated intensity, for example, Figures \ref{fig:LogInt} (A) and (B). We will now set $\lambda_0=0$ and refit the fractional Hawkes process to the data sets, to further investigate if $c^\beta$ does indeed have some constant physical interpretation.  Explicitly, the conditional intensity function is
\begin{equation} \label{eq:FHP0Lamb0}
    \lambda(t)=\alpha\sum_{t_i<t}e^{\gamma(M_i-M_0)}cf_\beta(c(t-t_i)).
\end{equation}

Performing parameter estimation using likelihood methods requires a work around. Since $\lim_{t\to t_1^-}\lambda(t)=0$, one cannot consider $t_1$ to be an event, as otherwise the log-likelihood function is undefined. Instead, for events $\{t_i,M_i\}_{1\leq i\leq N}$ take the time interval to be $[t_1,t_N]$ and use $\{t_i,M_i\}_{2\leq i\leq N}$ as the data. 
 
 \begin{table}
 \caption{\label{tab:0Lamb0FHPestimates}MLEs, the negative log-likelihood and the AIC for the restricted Fractional Hawkes Process.}
 \centering
\begin{tabular}
{ |p{1.7cm} p{1cm} p{1cm} p{1cm} p{1cm} p{1cm} p{1.3cm} p{1.3cm}| }
  \hline

 Catalogue&$\hat \alpha$&$\hat \gamma$ & $\hat \beta$& $\hat c$ &$\hat{c}^{\hat \beta}$&$-\ell(\hat{\bm \theta})$& AIC\\

JT &0.511 &1.228 & 0.518 &4.326 &2.137&-2298.4&-4588.9 \\

LM &0.435&1.266&0.608&0.660&0.777&-3319.4&-6630.9\\

HM &0.292 &1.455 &0.652&0.913&0.943&-3073.8&-6139.5\\
 
\hline
\end{tabular}

\end{table}
 
We again perform the graphical residual check for the restricted fractional Hawkes process, and construct the most extreme values of $c^\beta$ using the asymptotic confidence intervals.

\begin{figure}[!ht]
    \centering
    \includegraphics[width=0.65\textwidth]{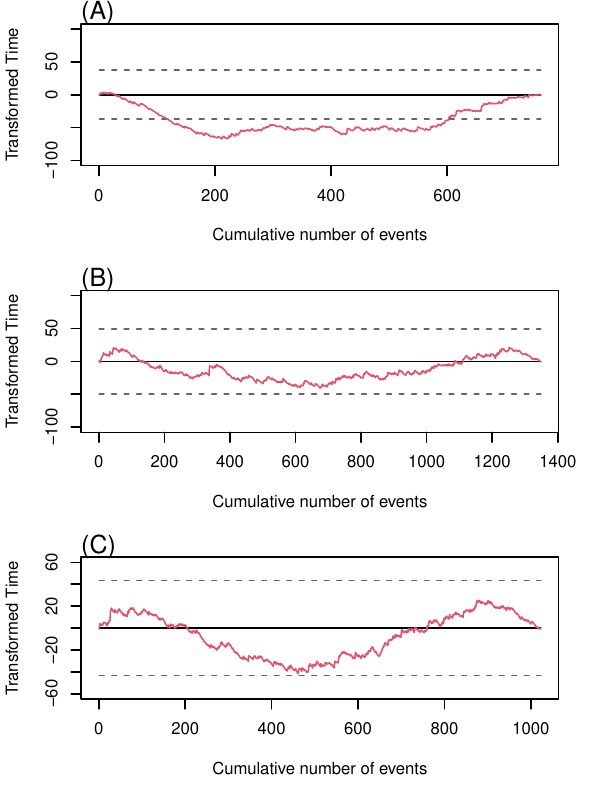}
    \caption{Theoretical mean removed transformed time for the restricted fractional Hawkes process. (A) corresponds to the ``JT", (B) the ``LM" and (C) ``HM" data sets respectively. The dashed lines correspond to the 95\% confidence intervals.}
    \label{fig:0LambResid}
\end{figure}

The graphical residual check for serial correlation, Figure \ref{fig:0LambCorrelation} in the supplemental material, indicates there is no major change between $U_k$ and $U_{k+1}$, as in, there once again there is no concern of correlation. Moreover, the transformed time check suggests that the restricted fractional Hawkes process describes most features of the catalogues ``LM" and ``HM" while still failing to model the delayed onset quiescence of the ``JT" catalogue. This is somewhat to be expected since once can consider this the fractional Hawkes process with $\lambda_0=0$ to be nested within the fractional Hawkes process with $\lambda_0 \geq 0$. Intuitively, a simpler model will not be able to explain more features of data than the full model.

Asymptotic confidence intervals at a 95\% and 99\% level for $\hat c$ and $\hat \beta$ are now constructed and the most extreme values of $\hat c^{\hat \beta}$ are reported in Table \ref{tab:cbetaCI2}.
 
\begin{table}
\caption{\label{tab:cbetaCI2} Most extreme values of $c^\beta$ using likelihood confidence intervals for the restricted confidence intervals.}
\centering
\begin{tabular}{ |p{1.7cm} p{1.7cm} p{1.7cm} p{1.7cm} p{1.7cm}| }

 \hline
 Catalogue& Min 95\%&Max 95\%&Min 99\%& Max 99\% \\

JT &1.882&2.432&1.810&2.535\\

LM &0.376&1.026&0.225&1.125\\

HM &0.655&1.189&0.547&1.271\\ \hline

\end{tabular}

\end{table}
 
It would appear that the true value of $c^\beta$ could overlap for the ``LM" and ``HM" data sets. As previously, however, the potential value for the ``JT" data set appears to differ. This may further suggest that the value of $c^\beta$ depends on the stage of the seismic cycle and geological setting, however since the restricted fractional Hawkes process fails the residual check this conclusion is weak.

To briefly compare the AIC values in Tables \ref{tab:AICFHPvETAS} and \ref{tab:0Lamb0FHPestimates}, in all cases the full fractional Hawkes process outperforms the restricted version (AIC in favour of full model with difference greater than 2), and the ETAS model outperforms the restricted fractional Hawkes process on the ``JT" and ``HM" data sets. This would suggest there is a non-negligible background rate of events during the study period. The restricted fractional Hawkes process however outperforms the ETAS model on the ``LM" data set. Since for this catalogue $\tau_N \approx N$ then the difference in log-likelihoods is due to the sum term of equation \eqref{eq:loglik}. This would suggest the log intensity function for the restricted fractional Hawkes process is generally greater than that of the ETAS model at event times. Therefore, further demonstrating the advantage of the unbounded kernel function of the fractional Hawkes process.
\section{Conclusion}\label{sec:Conc}
The fractional Hawkes process is a new point process model for earthquakes that has both empirical and theoretical backing. A maximum likelihood estimation scheme for the fractional Hawkes process was developed and its consistency was confirmed by a simulation study. Fitting this model to three mainshock-aftershock sequences from the Southern California seismological zone, as well as a sequence from New Zealand, and comparing it to the ETAS model suggests that in the right scenario the fractional Hawkes process can outperform the ETAS model. Comparing to the ETAS model the fractional Hawkes process best models earthquake sequences with stronger clustering, and with a relatively low cut-off magnitude, both of which require relatively complete data. This suggests that when developing a regime switching type of point process model for long-term seismicity, such as Markov modulated point processes with an example being that of \cite{Wang2012}, one can consider using the fractional Hawkes process for the periods with more clustered aftershocks and use the ETAS model for the less clustered periods. 

Additionally, in the restricted geological setting studied it was found that the value of $\hat c^{\hat \beta}$ may have a physical interpretation as a constant dependent on the geology and stage of the seismic cycle. This requires confirmation across more geological settings and stages of the seismic cycle, albeit there is some promise for such a result.

The true strength of the fractional Hawkes process will be seen when it is extended. Extensions of the fractional Hawkes process should be analytically tractable without any major simplifying assumptions, which is a downfall of the ETAS model and existing Markov-modulated point process models. Briefly, the ETAS model can only forecast short-term aftershock sequences \citep[e.g.][]{harte2013bias} and is generally intractable. Current regime switching models are also somewhat limited. In order to be tractable they are regularly limited to a memoryless continuous time Markov chain \citep[e.g.][]{Langrock2013MMPP} or a piecewise constant occurrence rate \citep[e.g.][]{Wang2012}. The Mittag-Leffler function's closed form Laplace transform and strong connection to the vast theory of fractional calculus should allow these assumptions to be relaxed in an extended fractional Hawkes process. Such a model should allow for feedback from the event history, and so long-range forecasts will depend on previous event times and magnitudes hopefully improving both their usability and sensibility. This is the goal of an extended model currently being developed. 

\section{Acknowledgements}
The authors wish to acknowledge the use of New Zealand eScience Infrastructure (NeSI) high performance computing facilities, consulting support and/or training services as part of this research. New Zealand's national facilities are provided by NeSI and funded jointly by NeSI's collaborator institutions and through the Ministry of Business, Innovation \& Employment's Research Infrastructure programme. URL https://www.nesi.org.nz.

\bibliographystyle{chicago}
\bibliography{BIBforPaper}
\clearpage
\appendix
\section{Supplementary Figures}
\begin{figure}[!ht]
    \centering
    \includegraphics[width=\textwidth]{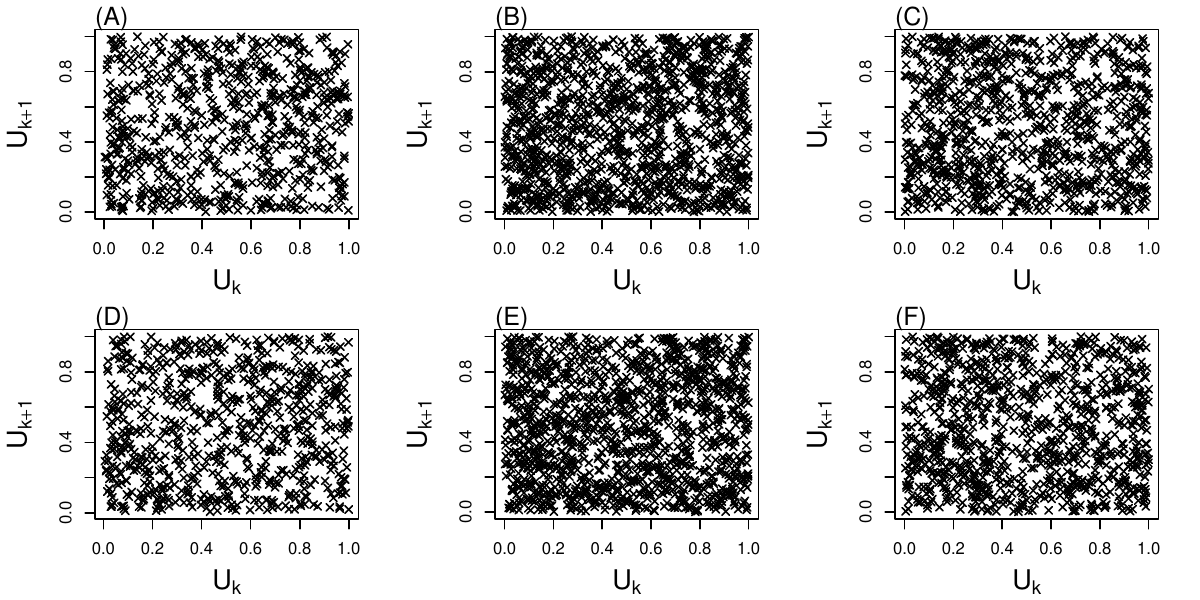}
    \caption{Transformed time difference residual check for serial correlation. (A), (B) and (C) are fits of the ETAS model in order ``JT", ``LM" and ``HM". (D), (E) and (F) are fits of the fractional Hawkes process for the same order of data sets.}
    \label{fig:TotalCorrelation}
\end{figure}
\begin{figure}[!ht]
    \centering
    \includegraphics[width=\textwidth]{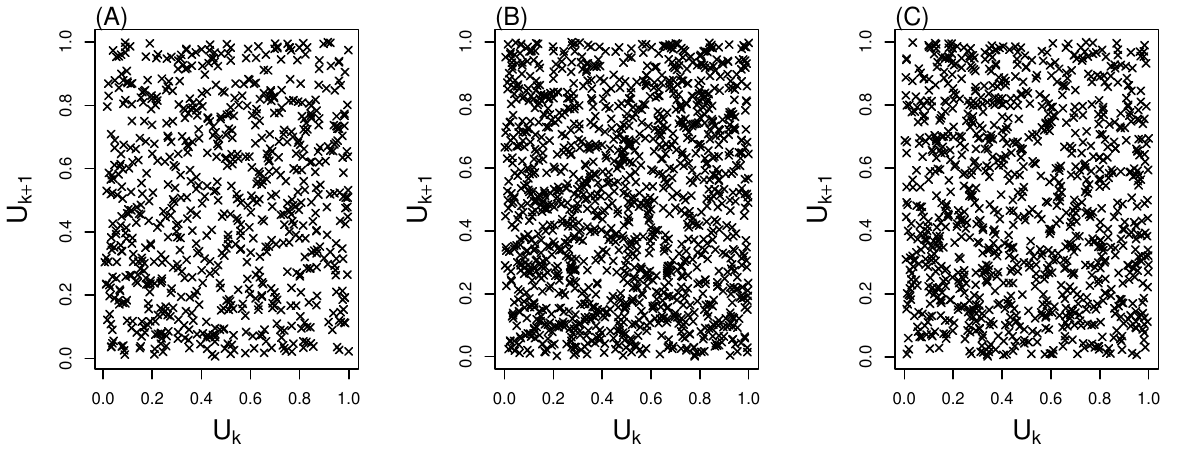}
    \caption{Graphical serial correlation check for the restricted fractional Hawkes process. (A), (B) and (C) correspond to the ``JT", ``LM" and ``HM" data sets respectively.}
    \label{fig:0LambCorrelation}
\end{figure}

\end{document}